\documentclass[12pt]{article}
\usepackage[style=authoryear,backend=bibtex]{biblatex}
\usepackage[parfill]{parskip}
\usepackage{multicol}
\usepackage{amsthm}
\usepackage{subcaption}
\usepackage{amsmath,times,hyperref}
\usepackage{amssymb}
\usepackage{amsfonts}
\usepackage{latexsym}
\usepackage{graphicx}
\usepackage{bbm}
\usepackage{mathtools}
\usepackage{bm}
\usepackage{algorithm}
\usepackage[usenames,dvipsnames]{color} 
\usepackage{centernot}
\usepackage{ifthen}
\usepackage{setspace}
\usepackage{tabu}
\usepackage{framed}
\usepackage[dvipsnames]{xcolor}
\usepackage[margin=1in]{geometry}
\usepackage[symbol]{footmisc}

\begin{document} 

\begin{center}
    {\large
    Pricing the COVID-19 Vaccine: A Mathematical Approach\footnote[1]{Preprint submitted on 29 Dec. 2020}}
    
    \vspace{5mm}
    
    \begin{multicols}{3}
    Susan Martonosi$^{a,}$\footnote[2]{Corresponding author}\\
    {\small martonosi@g.hmc.edu}
    
    Banafsheh Behzad$^b$\\
    {\small Banafsheh.Behzad@csulb.edu}
    
    Kayla Cummings$^c$\\
    {\small kaylac@mit.edu}
    
    \end{multicols}
    
    {\it \small   $^a$Department of Mathematics, Harvey Mudd College\\
    $^b$Department of Information Systems, College of Business Administration at California State University, Long Beach\\
    $^c$Operations Research Center, Massachusetts Institute of Technology}
\end{center}

\vspace{10mm}

\hrule 

{\bf Abstract}

According to the World Health Organization, development of the COVID-19 vaccine is occurring in record time. Administration of the vaccine has started the same year as the declaration of the COVID-19 pandemic. The United Nations emphasized the importance of providing COVID-19 vaccines as “a global public good”, which is accessible and affordable world-wide. Pricing the COVID-19 vaccines is a controversial topic. We use optimization and game theoretic approaches to model the COVID-19 U.S. vaccine market as a duopoly with two manufacturers Pfizer-BioNTech and Moderna. The  results  suggest  that  even  in  the  context  of  very  high production and distribution costs, the government can negotiate prices with the manufacturers to keep public sector prices as low as possible while meeting demand and ensuring each manufacturer earns a target profit. Furthermore, these prices are consistent with those currently predicted in the media.

{\it \footnotesize Keywords: coronavirus, COVID-19, healthcare, optimization, game theory, public policy, vaccine pricing} 

\medskip
\hrule

\section{Introduction} \label{s:intro}

The first case of Coronavirus Disease 2019 (COVID-19) 
was recognized in Wuhan, China, in December 2019 and was then announced as pandemic by the World Health Organization (WHO) on March 11th, 2020 (\cite{cdccovid}). This pandemic has become one of the world's main public health challenges due to both transmission speed and scale, and the fact that certain segments of the population 
are at increased risk of severe illness and death (\cite{DOSSAN}). As of early December 2020, there have been more than 66M confirmed cases and 1.5M deaths worldwide due to COVID-19 (\cite{WHOt}). 

Vaccine development is a complicated process normally requiring years of research and development. There are several ongoing efforts around the world to discover a vaccine that mitigates COVID-19 transmission. 
New-generation vaccines including messenger RNA (mRNA) vaccines are highly effective and have potential  for fast development, low cost manufacturing and safe administration. Thus, these vaccines can serve as an assuring substitute for conventional vaccines (\cite{mrna}). As of Dec. 5, 2020, the Coronavirus Vaccine Tracker by New York Times (\cite{nytvt}) reports that there are 58 COVID-19 vaccines in clinical trials on humans, of which there are two leading manufacturers in the U.S. in Phase 3 of developing mRNA vaccines: Pfizer-BioNTech and Moderna. The collaboration between Pfizer and the German company BioNTech reports their vaccine to be 95\% effective in preventing COVID-19 (\cite{Pfizer95}), and their vaccine was approved for emergency use in the U.K. on December 2, 2020 (\cite{NYT_PF95}).  (Throughout this work, we refer to the Pfizer-BioNTech collaboration as simply ``Pfizer'').  In partnership with National Institute of Health (NIH), Moderna began vaccine development in January 2020; 
on Nov. 16, 2020, Moderna announced its vaccine is 94.5 percent effective, joining Pfizer on the frontier of the worldwide endeavor to develop a COVID-19 vaccine (\cite{moderna}).  

On July 22, 2020, Pfizer announced an agreement with the U.S. government for an initial order of 100 million doses of its mRNA vaccine for \$1.95 billion with a possibility of acquiring up to 500 million additional doses (\cite{pa}). On August 11, 2020, Moderna, Inc., announced that the  U.S. government committed up to \$2.48 billion for early access to 100 million doses of Moderna's mRNA vaccine, with the option to purchase up to 400 million additional doses (\cite{ma}). 
According to an article published on Nov. 17, 2020, Pfizer has set the initial price at \$19.50 a dose (\$39 per patient), and Moderna has set its vaccine price to \$25 per dose (\$50 per patient) (\cite{forbesprice}). \cite{KFF} report the price of the Moderna vaccine as \$30 for the two-dose regimen.

Distribution of the mRNA vaccines is a serious challenge due to the need for ultra-cold storage. The Pfizer and Moderna vaccines require different storage temperatures: $-70$ degrees Celsius for Pfizer and $-20$ degrees Celsius for Moderna (\cite{npr}). 
Moderna's vaccine can be stored in a regular freezer, making its distribution less costly. Pfizer has provided several ultra cold freezers to store and distribute their own vaccine (\cite{CT}).

The Centers for Disease Control and Prevention (CDC) is the primary public health organization in the U.S. charged with the responsibility of development and fulfillment of vaccines and ensuring adequate societal immunization coverage levels. The Centers for Medicare and Medicaid Services released a comprehensive plan to ensure all Americans have access to the COVID-19 vaccine at no cost once available (\cite{cms}). According to \cite{KFF}, the COVID-19 vaccine should be free to everyone in the U.S. under the CDC's COVID-19 Vaccination Program (\cite{covidvacpr}). The CDC discusses vaccine coverage under five program types: Medicare, Vaccine for Children (VFC), Medicaid, Section 317 (Vaccines for uninsured adults),  the Veterans Administration, and private insurance. 
We characterize the U.S. COVID-19 vaccine as a two-sector market, similar to other vaccine markets. The private sector includes the vaccines that are purchased by private insurers, and the public sector includes all other governmental programs. Typically public sector vaccine costs are lower. 


Several studies in the operations research literature use theoretical and empirical approaches to analyze multiple aspects of the COVID-19 pandemic. These aspects include modeling the spread of the disease (\cite{Ivorra2020}), ICU capacity management (\cite{Alban2020}), importance of wearing high quality masks (\cite{Rosenstrom2020}), ventilator inventory management (\cite{Mehrotra2020}), the impact of COVID-19 on global supply chain performance and food supply chain (\cite{Ivanov2020} and \cite{Singh2020}), and plasma donations during the COVID-19 pandemic (\cite{Nagurney2020}), among others. To the best of our knowledge, there are no theoretical or empirical studies on COVID-19 vaccine pricing. This study fills this gap.

Previous work by \cite{CummingsEtAl2020} models the negotiation of public sector pediatric vaccine prices by the CDC in the context of a duopoly.  The authors develop an optimization model for minimizing public sector expenditures while ensuring adequate public sector vaccine supply and adequate profit margins for each manufacturer (preventing a monopoly). They treat the private sector market as an exogenous duopoly price competition operating under the surplus conditions of a Bertrand-Edgeworth-Chamberlin (BEC) equilibrium, as derived in \cite{bj}.  

This paper considers the U.S. COVID-19 vaccine market as a duopoly between Pfizer and Moderna. We adapt the model of \cite{CummingsEtAl2020} to determine prices for the U.S. public sector COVID-19 vaccines
. The analysis suggests that even in the context of high production and distribution costs, the U.S. government can negotiate prices with the manufacturers that keep public sector prices as low as possible while meeting demand and ensuring each manufacturer earns a target profit.  Moreover, these prices are consistent with those currently predicted in the media.  Prices are likely to decrease over time as the cold chain becomes more efficient.  Finally, the model presented here can be adapted to the context of an oligopoly when new vaccines enter the market.

Section \ref{s:model} presents our adaptation of the model to the COVID-19 vaccine scenario.  In Section \ref{s:data} we justify the parameter values used for our empirical tests.  Section \ref{s:analysis} presents the analysis and discussion.  
Section \ref{s:conclusion} concludes.

\section{Model} \label{s:model}

We use the same notation and largely the same model as in \cite{CummingsEtAl2020}.  The sets, variables and parameters used in the optimization model can be found in Table \ref{T:modelComponents}.  The full optimization model is given in Figure \ref{F:model}, which we describe now.

\begin{table}[htbp!]
\begin{center}
\small
{\tabulinesep=1mm
\begin{tabu}{|l|l|p{0.35\linewidth}||p{0.35\linewidth}|}
\hline
Name & Type & Meaning & Estimated Value(s) \\ \hline  \hline
$M$ & set & manufacturers &   \\ 
$S$ & set & sectors \{\text{pub}, \text{priv}\} &  \\ \hline
$p_{i}^{\text{pub}}$ & variable & public sector vaccine price, $i \in M$ &  \\
$q_{i}^{\text{pub}}$ & variable & public sector vaccine quantity, $i \in M$ &  \\
$z$ & variable & manufacturers' absolute public sector price difference &  \\
\hline
$p_i^{\text{priv}}$ & parameter & private sector vaccine price, $i \in M$ & Eq. (\ref{E:BECeqPrice})  \\
$q_i^{\text{priv}}$ & parameter & private sector vaccine quantity, $i \in M$ & Eq. (\ref{E:BECeqQuant})   \\
$D$ & parameter & total U.S. demand & [157.05M, 174.5 M, 191.95M] \\
$a^s,\, b,\, c$ & parameters & demand curve coefficients, $s \in S$ & Eqs. (\ref{M:interceptCov}), (\ref{E:b}), (\ref{E:c})  \\
$k$ & parameter & order of magnitude difference between prices and quantities per manufacturer and sector$^{\dagger}$ & $6$  \\
$U$ & parameter & private sector capacity lower bound for surplus & Eq. (\ref{E:Ucalc})  \\
$K_i$ & parameter & total production capacity, $i \in M$ & $K_{Pf}=250$M, $K_{Mod}=200$M  \\
$P_i$ & parameter & minimum profit threshold, $i \in M$ &  \parbox[t]{\linewidth}{ $P_{Pf}: \left[\$25.7M, \$234M, \$2.57B\right]$\\$P_{Mod}: \left[\$41.4M, \$496M, \$2.57B\right]$}\\
$d_i$ & parameter & two-dose production and distribution cost, $i \in M$ $^{\dagger}$&  $\left[\$0, \$6.6, \$23.44, \$31.96\right]$  \\
$\gamma$ & parameter & product similarity &  [0.25, 0.50, 0.75]  \\
$r^s$ & parameter & \% total demand sold in sector $s \in S$ $^{\dagger}$ & $r^{\text{pub}} = 0.57$  \\ 
$\mu$ & parameter & 
objective function weight$^{\ddagger}$  & 0.9  \\ 
\hline
\end{tabu}}
\caption{Components of the optimization model in Figure \ref{F:model}, and their estimated base value and range for sensitivity analysis.  Quantities, prices and costs reflect two-dose regimens of the vaccine.  \\ $^\dagger$: new parameters not used in \cite{CummingsEtAl2020}$^1$. \\ $^\ddagger$: For numerical stability, all vaccine quantities are scaled to units of one million in the optimization model. \\
$^1$\cite{CummingsEtAl2020} include a parameter $\rho$ representing a legislative limit on year-over-year price increases. No such legislation has yet been introduced for the COVID-19 vaccine (\cite{KFF}), so we omit this requirement.}
\label{T:modelComponents}
\end{center}

\end{table}

\begin{figure}[htbp!]
\small
{ \begin{equation}\label{E:objectiveFunction}
\text{minimize }\quad \mu\Big(\sum_{i \in M}q_{i}^{\text{pub}}\cdot p_{i}^{\text{pub}}\Big) + (1-\mu) z
\end{equation}
\begin{align}
\sum_{i \in M}q_i^{\text{pub}} &\geq r^{\text{pub}} D  \label{E:pubDemand} \\ 
q^s_i + bp^s_i - cp^s_j &= a^s && i, j \in M, \, i \neq j, s \in S \label{E:demCurve} \\  
\sum_{s \in S} q^s_i(p^s_i-d^s_i) &\geq P_i && i \in M \label{E:targetProfit} \\ 
\sum_{s \in S} q_i^{s}  &\leq K_i && i \in M \label{E:capacity} \\ 
K_i - q_i^{\text{pub}} &\geq U && i \in M  \label{E:surplus} \\ 
\qquad \qquad \qquad z & \geq p_i^{\text{pub}}- p_j^{\text{pub}} &&i,j \in M,\, i \neq j \label{E:absval} \\ 
q_i^{\text{pub}},\, p_i^{\text{pub}},\, z &\geq 0 && i \in M  
\end{align}}
\caption{{ Model to optimize public-sector price-setting dynamics in a surplus scenario. 
}} 
\label{F:model}
\end{figure}

We consider two manufacturers $(i \in M)$ of competing and differentiated vaccines.  The manufacturers sell these vaccines in two sectors, $s \in S = \{\text{pub}, \text{priv}\}$.  When selling in the private sector the prices, $p^{\text{priv}}_i$, are determined in a Bertrand-Edgeworth-Chamberlin (BEC) duopoly game presented in \cite{bj}. We also assume a surplus regime, in which each manufacturer's production capacity, $K_i$, for the new vaccine is sufficiently large that each can maximize private-sector profits unconstrained by capacity.  

Within the public sector, the CDC negotiates quantities, $q^{\text{pub}}_i$, and prices, $p^{\text{pub}}_i$, with each manufacturer  to minimize the total government cost while ensuring public sector demand can be fully met. To reflect the fact that competing vaccines tend to have similar prices, the CDC chooses among cost-minimizing solutions those which also minimize the absolute public sector price difference between the two manufacturers, $z = |p^{\text{pub}}_1-p^{\text{pub}}_2|$, yielding the weighted objective function in Equation (\ref{E:objectiveFunction}).  

The negotiated prices and quantities are constrained by the following considerations:

\begin{itemize}
    \item Public sector demand must be met (Eq. \ref{E:pubDemand}).  We introduce parameter $r^{s}$, which is the percentage of total demand $D$ attributed to sector $s$, such that $r^{\text{pub}} + r^{\text{priv}} = 1$.  
    
    \item Prices and quantities follow linear demand curves in each sector (Eq. \ref{E:demCurve}).
    
    \item Each manufacturer $i \in M$ must achieve a target profit $P_i$ (Eq. \ref{E:targetProfit}).  As a novel departure from the model of \cite{CummingsEtAl2020}, we introduce per-unit production and distribution costs $d_i$, reflecting the significant expense of the cold chain required to distribute mRNA vaccines. 
    
    \item Each manufacturer's total quantity sold cannot exceed their total capacity (Eq. \ref{E:capacity}).
    
    \item Each manufacturer's residual private sector capacity is determined by public sector quantities and must exceed the bound $U$ governing the BEC game's surplus equilibrium regime (Eq. \ref{E:surplus}). 
    
    \item The absolute price difference, $z$, between the two manufacturers is expressed as linear inequality bounds that the objective function drives to tighten (Eq. \ref{E:absval}).
\end{itemize}

The private sector prices and quantities are determined exogenously by the equilibrium solution to the BEC game given in \cite{CummingsEtAl2020}:   \begin{equation} \label{E:BECeqPrice} p_i^{\text{priv}} = \frac{a^{\text{priv}}}{2b-c}, \forall i \in M. \end{equation}
\begin{equation} \label{E:BECeqQuant} q_i^{\text{priv}} = bp_i^{\text{priv}}, \forall i \in M, \end{equation}
\noindent where \begin{equation}\label{E:b} b =  \frac{10^k}{(1+\gamma)(1-\gamma)}, \end{equation}
\noindent (for $k$ the approximate order of magnitude difference between prices and quantities produced by each manufacturer in each sector), \begin{equation}\label{E:c} c = \gamma b, \end{equation}
\noindent and $a^{\text{priv}}$ is estimated as described in Section \ref{ss:demcurveparams}. The capacity bound $U$ is given by: \begin{equation} \label{E:Ucalc} U = \frac{a^{\text{priv}}(1+\gamma)}{\gamma} \cdot \left(1-\frac{2(1-\gamma)^{0.5}}{(1+\gamma)^{0.5}(2-\gamma)}\right). \end{equation}

Note that Pfizer and Moderna are both anticipated to produce vaccines requiring a two-dose regimen.  We assume all individuals will receive both doses, so we interpret quantities, unit prices, and unit costs in terms of individuals served in two-dose increments.

\section{Parameter Estimation} \label{s:data}

Here, we discuss estimates and sensitivity analysis ranges for the model parameters describing the Pfizer and Moderna vaccines, summarized in the last  column of Table \ref{T:modelComponents}.

\subsection{Demand Curve Parameters} \label{ss:demcurveparams}

We estimate the linear demand curve parameters $a^{\text{pub}}, a^{\text{priv}}, b,$ and $c$ in the same manner as \cite{CummingsEtAl2020}. This method uses publicly available demand and price data to determine the average public and private sector intercepts, $a^{\text{pub}}$ and $a^{\text{priv}}$. 
Historical demand and price data are not yet available for the COVID-19 vaccine, so we use data available for the influenza (flu) vaccine as a proxy.   The flu vaccine is the most widely distributed vaccine in the U.S., with approximately 175 million U.S. individuals receiving the vaccine in the 2019-20 contract year (\cite{CDCfludoses}).  This is the same order of magnitude as the number of U.S. individuals who could receive the COVID-19 vaccine during its first year of production.  In the U.S. flu vaccine market two primary manufacturers market vaccines to people older than six months of age: Sanofi Pasteur (Fluzone\textregistered Quadrivalent) and GlaxoSmithKline (Fluarix\textregistered Quadrivalent).  Table \ref{T:demEst} summarizes public and private sector prices for these two vaccines.

The new COVID-19 vaccines will require two doses per person, as compared to the flu vaccine's single dose per person.  Additionally, the COVID-19 vaccines are expected to have much higher distribution costs relative to the flu vaccine.  Thus to estimate $a^s$ in each sector, we use the reported number of administered doses of flu vaccine from the years 2010-11 through 2019-20 to reflect the number of individuals who will seek the COVID-19 vaccine.  We double the prices given in Table \ref{T:demEst} to reflect the cost of a two-dose regimen, and we add the estimated two-dose production and distribution costs $d_i$ that will naturally raise prices. 
Because two-dose quantities in each sector by each manufacturer are likely to be on the order of tens of millions, and prices are likely to be on the order of tens, we set $k=6$. Incorporating these modifications into the approach of 
\cite{CummingsEtAl2020}, we obtain the following expression for the estimated linear demand curve intercept:

\begin{equation}\label{M:interceptCov}
a^s = \frac{1}{Y}\sum_{y=1}^Y \left(\frac{1}{2}\sum_{i \in M}q_{iy}^{s} + \frac{10^6}{2+2\gamma}\sum_{i \in M}(2p_{iy}^{s}+d_{iy}^s) \right) \qquad s \in S,
\end{equation}
\noindent where $q_{iy}^s$ and $p_{iy}^s$ are, respectively, the quantity of flu vaccine sold and the price set by manufacturer $i \in M$ in sector $s \in S$ during year $y \in \{1, \dots, Y\}$.  We can interpret $a^s$ to be the number of sector $s$ individuals who would seek a two-dose regimen of the COVID-19 vaccine from either manufacturer if the vaccine had zero cost.  

\begin{table}[htbp!]
\small
\begin{center}
{\tabulinesep=1mm
\begin{tabu}{|l||c|c|c|c|c|}
\hline
& \multicolumn{2}{c|}{Public Sector} & \multicolumn{2}{c|}{Private Sector} & \\
\hline

Year & Sanofi Pasteur & GlaxoSmithKline & Sanofi Pasteur & GlaxoSmithKline  & Total Demand (M)\\
\hline \hline
2010-11	&	\$10.64 	&	\$8.90 	&	\$13.16 	&	\$10.98 	&	155.1	\\
2011-12	&	\$11.68 	&	\$10.97 	&	\$13.16 	&	\$12.41 	&	132.0	\\
2012-13	&	\$11.68 	&	\$9.25 	&	\$14.56 	&	\$10.98 	&	134.9	\\
2013-14	&	\$17.05 	&	\$13.65 	&	\$20.50 	&	\$15.90 	&	134.5	\\
2014-15	&	\$17.44 	&	\$13.65 	&	\$21.09 	&	\$15.90 	&	147.8	\\
2015-16	&	\$17.94 	&	\$14.05 	&	\$21.70 	&	\$16.05 	&	146.4	\\
2016-17	&	\$19.14 	&	\$14.43 	&	\$23.17 	&	\$16.82 	&	145.9	\\
2017-18	&	\$15.68 	&	\$14.43 	&	\$18.72 	&	\$16.82 	&	155.3	\\
2018-19	&	\$15.11 	&	\$13.50 	&	\$19.26 	&	\$16.82 	&	169.1	\\
2019-20	&	\$13.76 	&	\$13.50 	&	\$18.31 	&	\$16.82 	&	174.5	\\

\hline 
\end{tabu}}
\end{center}
\caption{Yearly flu vaccine prices per dose, by sector (\cite{CDCfluprices}), and total demand in millions of doses (\cite{CDCfludoses}).
}
\label{T:demEst}
\end{table}

We use the 2019-20 total flu vaccine demand of 174.5M as the baseline two-dose demand, $D$, considered by our model, and consider values 10\% below  and above that baseline.

\subsection{Production Parameters}

The two COVID-19 vaccines are still in development at the time of this writing, so we rely on information contained in press releases and other media to estimate  production-specific parameters.    

\subsubsection{Total Production Capacity, $K_i$.}
As stated in Section \ref{s:intro}, Pfizer and Moderna have the ability to produce up to 500 million and 400 million doses, respectively, for the U.S. market in 2021. We use $K_{Pf}=250M$ and $K_{Mod}=200M$ as the capacity measured in two-dose regimens. Moreover, Pfizer and Moderna have been rapidly ramping up their supply chains, and it is expected that both manufacturers will have considerably more capacity than this in 2021, permitting us to safely assume the surplus regime of the BEC equilibrium.  For instance, Pfizer plans to produce 1.3B doses in 2021 for global use (\cite{CT}).  

\subsubsection{Minimum Profit Threshold, $P_i$.}
Articles in the media question whether Pfizer and Moderna are trying to attain a profit with the new COVID-19 vaccine (\cite{healthline, marketplace}).  The U.S. government has an incentive for the two manufacturers to make a profit on this vaccine, as this stabilizes the market  and also ensures that pharmaceutical companies will be inclined to once again rapidly develop a vaccine if there is another global pandemic in the future.  

We develop three estimates for each manufacturer's target profit
.  First, we divide each company's reported annual research and development (R\&D) budget by its reported number of products as an estimate of the average profit it hopes to recuperate from sales of the new vaccine.  Pfizer reports an annual R\&D budget of \$8.65B and 337 total pharmaceutical products, corresponding to a target profit of \$25.7M (\cite{PfFin,PfProd}).  However, many of these products are over-the-counter products or generic labels for other products they carry.  If we instead consider only the 37 products Pfizer lists as its primary products (\cite{PfProdKey}), we obtain a target profit of \$234M.  Moderna reports an annual R\&D budget of \$496.3M and 12 pharmaceutical products that have reached the clinical studies phase (none are yet approved), corresponding to a target profit of \$41.4M (\cite{ModFin, ModProd}). Because Moderna has not yet brought a product to market, its COVID-19 vaccine could reasonably be expected to recuperate the entire \$496.3M annual R\&D budget from the previous year, so we use this value as a second estimate of Moderna's target profit.  Finally, we can use prior year revenue estimates using the flu vaccine data reported in Table \ref{T:demEst} to obtain a third estimate of the companies' target profit, following \cite{CummingsEtAl2020}[S.M. Eq. 7]. In this way, we estimate a target profit of \$2.57B for each manufacturer.

\subsubsection{Two-Dose Production and Distribution Cost, $d_i$.}

Per-unit costs associated with the production and distribution of either manufacturer are not widely known.  We use estimates of per-dose investment costs for the flu vaccine provided by the World Health Organization (\cite{WHO})[Fig. 2].  In their figure, they report a mean per-dose cost of $\$3.30$ averaged over ten manufacturers and weighted by the quantities each produces.  We can also calculate an unweighted average of $\$15.98$ per dose, and an unweighted trimmed mean (discarding the smallest and largest reported investment costs) of $\$11.72.$  We again double these estimates, reflecting a two-dose regimen of the COVID-19 vaccine, to attain estimates for $d_i$ of $\$0$, $\$6.60$, $\$23.44$, and $\$31.96.$

\subsubsection{Other Parameters.}

The remaining parameters are the product similarity, $\gamma$, the public sector demand percentage, $r^{\text{pub}}$, and the objective function weight, $\mu$.  The degree of similarity between the two vaccine products depends on factors such as the mode of delivery and adverse effects.  In the absence of specific information about the vaccines, we test $\gamma = [0.25, 0.5, 0.75].$  
We assume that the public sector represents $r^{\text{pub}} = 57\%$ of the total U.S. vaccine market (\cite{odrh}). Lastly, with quantities scaled to millions, 
 $\mu=0.9$ effectively identifies, among cost-minimizing solutions, a solution for which the difference in public sector prices is minimal.




\section{Analysis and Discussion} \label{s:analysis}

We tested the 1,296 combinations of  model parameters specified in Table \ref{T:modelComponents} using the BARON solver on the NEOS server (\cite{baron, neos1, neos2, neos3}).  The model is infeasible whenever \textit{both} manufacturers have the highest target profit of \$2.57B, regardless of other parameters.  The vast majority of remaining infeasible cases occur when one manufacturer's target profit is at this highest value.  The high prices required to support such a high target profit erode demand to the point of failing to meet public sector demand.  Of the 554 feasible scenarios, 214 yield public sector prices and quantities that are nonzero for both manufacturers, reflecting the current expectation that both manufacturers will participate in the public sector market.  We discuss these 214 scenarios.

\subsection{Prices}

\begin{figure}[htbp!]
\small
    \centering
    \includegraphics[width=3.5in]{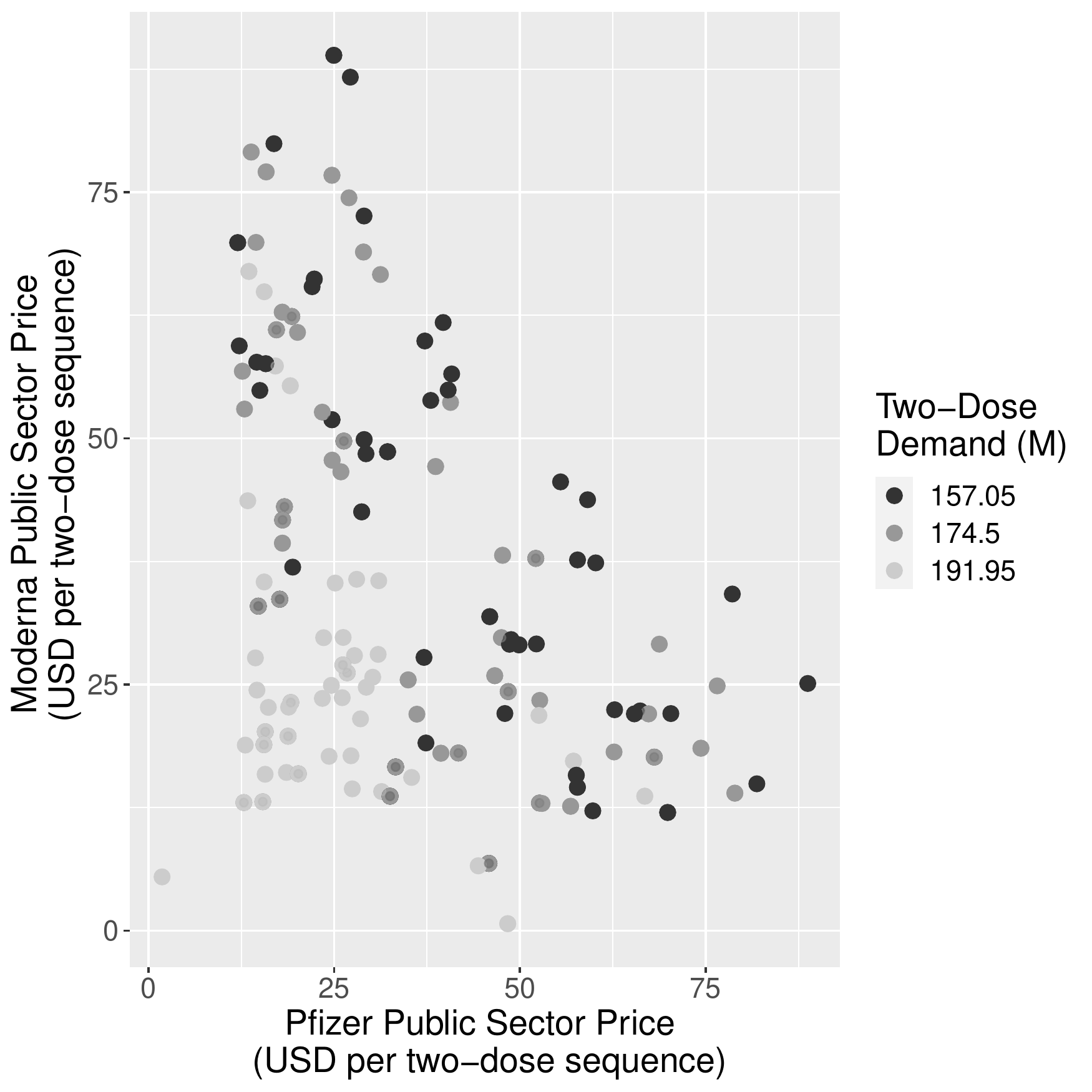}
    \caption{\footnotesize  Public sector prices (USD) determined by the optimization model for two-dose regimens for each manufacturer, as a function of total two-dose demand (M).
    }
    \label{F:pubpricesdem}
\end{figure}

Figure \ref{F:pubpricesdem} plots pairs of public sector prices of Moderna and Pfizer determined by the  model for each of the 214 scenarios, with shading determined by the total demand used in the scenario.  

First, we notice that the prices output by the model are reasonable and consistent with projections given in the news.  The overall centroid of the data points in Figure \ref{F:pubpricesdem} is located at a price of \$34.39 for Pfizer and \$35.09 for Moderna for a two-dose regimen.

Second, we detect three general bands of points, corresponding to the three values of total demand.  When total demand is projected to be low (157.05M two-dose regimens, in black), optimized prices are generally higher than when total demand is projected to be high (191.95M two-dose regimens, in light grey).  

Lastly, we notice a general inverse relationship in the prices of the two manufacturers: higher prices for Pfizer are generally coupled with lower prices for Moderna, and vice versa.  Extreme prices higher than \$75 per two-dose regimen occur when, e.g., both manufacturers have moderately high or high production costs and product similarity is low, meaning that the manufacturers have effectively segmented the market to their advantage.

\begin{figure}[htbp!]
\small
    \centering
    \begin{subfigure}[t]{0.4\textwidth}
        \includegraphics[height=6.5cm]{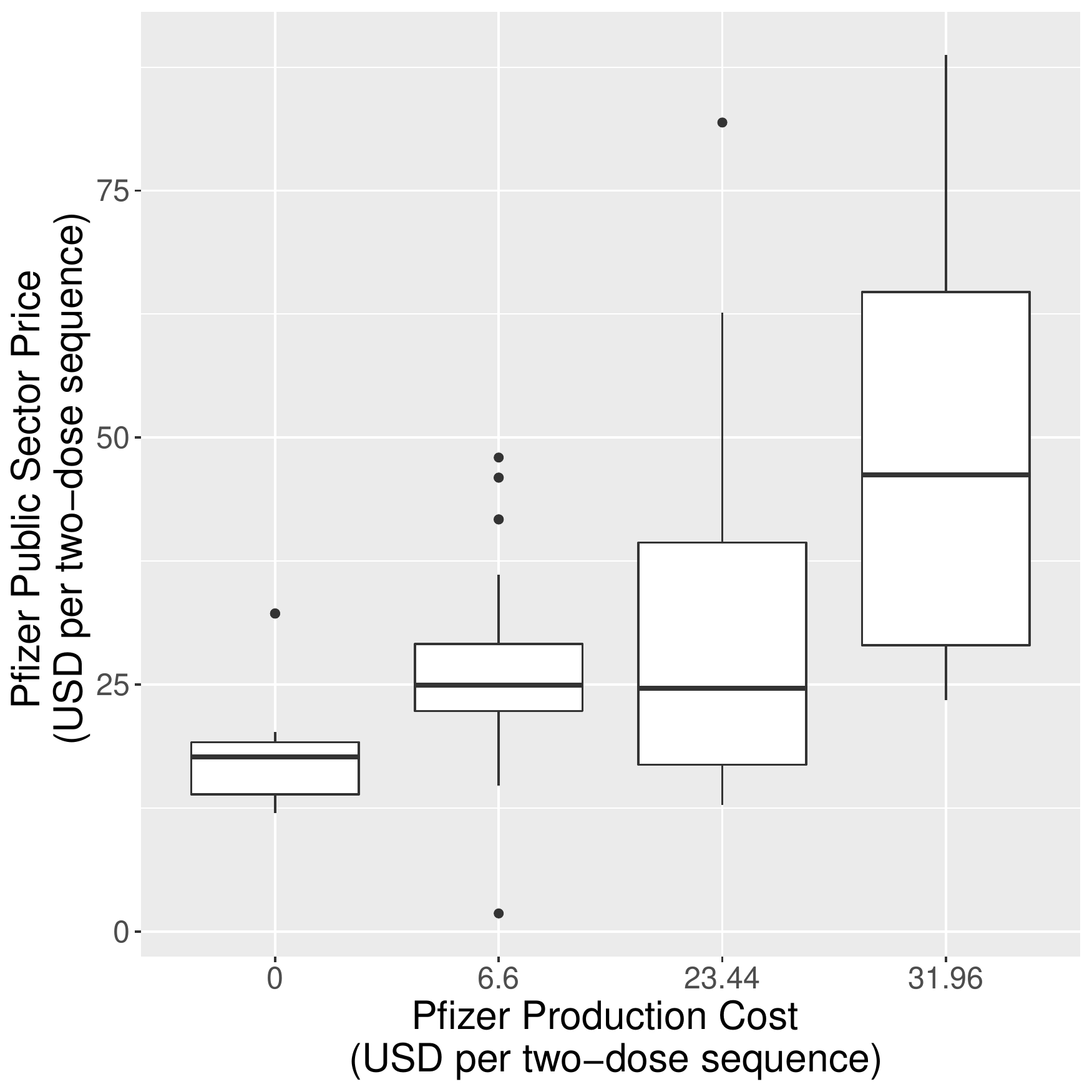}
        \caption{\footnotesize Pfizer public sector price distribution.}
        \label{F:Pfboxplot}
    \end{subfigure}
    ~
    \begin{subfigure}[t]{0.4\textwidth}
        \includegraphics[height=6.5cm]{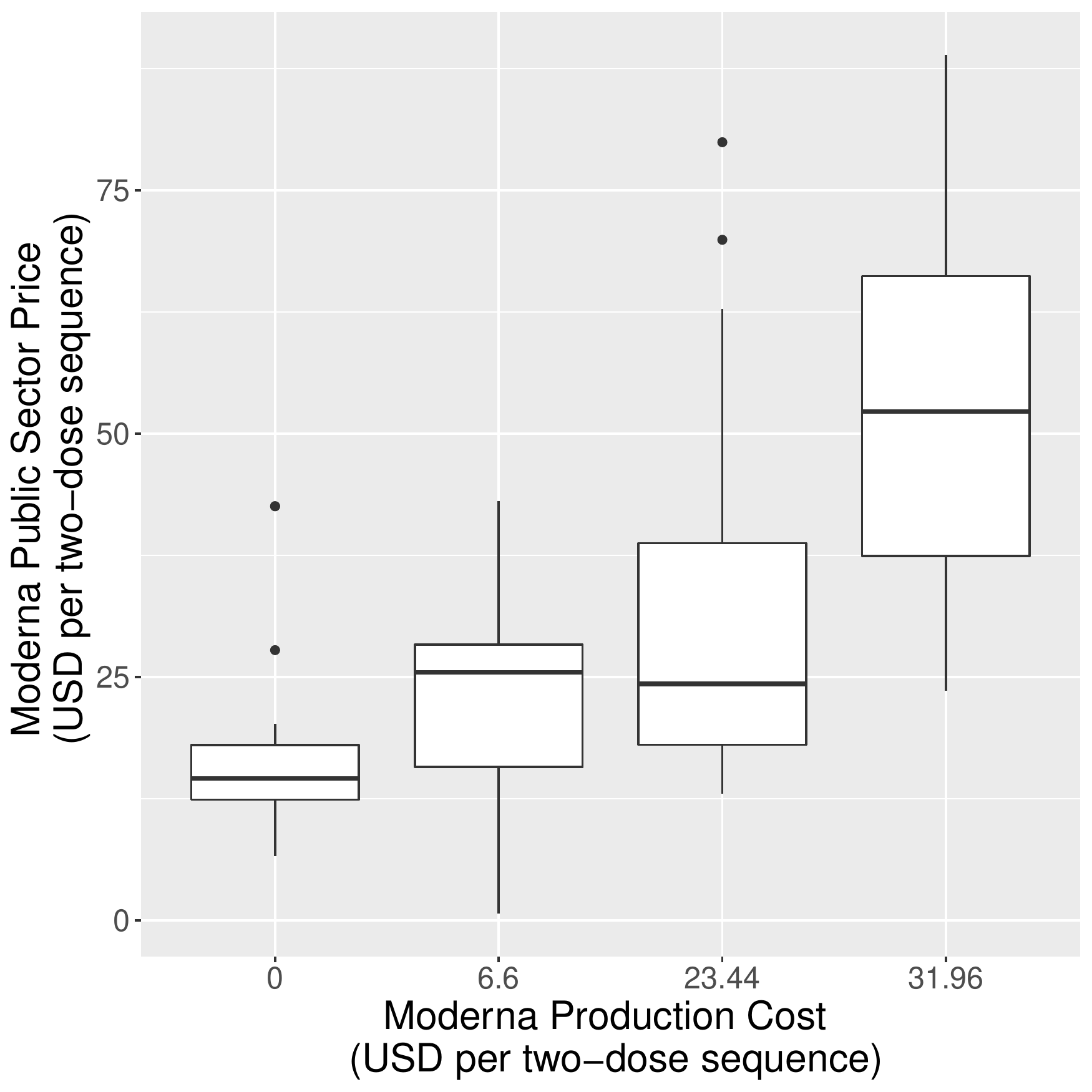}
        \caption{\footnotesize Moderna public sector price distribution.}
        \label{F:Modboxplot}
    \end{subfigure}
    \caption{Public sector price distributions determined by the optimization model for each manufacturer, by each manufacturer's production and distribution cost $d_i$.} \label{F:boxplotprodcost}
\end{figure}

Figure \ref{F:boxplotprodcost} plots the public sector price distribution of each manufacturer by that manufacturer's production and distribution cost, $d_i$.  We see that public sector prices generally rise with increasing production costs.  This is relevant because cold storage is quite costly  (\cite{Fortune}).  However, Moderna recently announced that its vaccine can tolerate a less stringent cold chain than initially thought, and it is hypothesized that Pfizer might eventually come to a similar conclusion.  Thus, we can expect  public sector prices to start high but eventually decrease 
as the temperature requirements relax, lowering distribution costs.

\subsection{Profit}

The CDC has an incentive to ensure minimum target profits are attained by both manufacturers.  This can prevent the formation of a monopoly caused by one manufacturer leaving the market.  Our model identifies public sector prices and quantities that guarantee each manufacturer achieves its target profit.

News articles have suggested possible prices of the upcoming vaccines by comparing the federal research funding offered to each manufacturer in exchange for a contracted number of doses if the vaccine is approved. For Pfizer, the federal support of \$1.95B in exchange for 100M doses indicates an effective public sector price of \$39 per two-dose regimen (\cite{pa}).  For Moderna, the federal support of \$2.48B in exchange for 100M doses amounts to a public sector price of \$49.6 per two-dose regimen (\cite{ma}).  

Using our model, we can identify scenarios in which both manufacturers secure a profit with public sector prices in the ranges cited.  Table \ref{T:rightprices} gives the settings of the tunable parameters from Table \ref{T:modelComponents} yielding public sector prices  in the range of $\$34-44$ per two doses for Pfizer and $\$45-55$ per two doses for Moderna.  Not only do the realized profits meet the projected target profits for each manufacturer in each of the five scenarios identified, realized profit for one of the two manufacturers substantially exceeds target profit in scenarios 1, 3, and 4.

\begin{table}[htbp!]
\small
\begin{center}
{\tabulinesep=1mm
\begin{tabu}{|l||c|c|c|c|c|}
\hline
Scenario	&	1	&	2	&	3	&	4	&	5	 \\
\hline \hline											
Total Demand, $D$ (M)	&	174.5	&	157.05	&	174.5	&	174.5	&	157.05	 \\
Product Similarity, $\gamma$	&	0.75	&	0.75	&	0.75	&	0.75	&	0.75	 \\
\hline											
Pfizer Target Profit, $P_{Pf}$ (\$M)	&	234	&	234	&	25.7	&	234	&	25.7	 \\
Pfizer Production Cost, $d_{Pf}$ (\$)	&	31.96	&	31.96	&	23.44	&	23.44	&	31.96	 \\
\hline											
Moderna Target Profit, $P_{Mod}$ (\$M)	&	41.4	&	41.4	&	41.4	&	41.4	&	496	 \\
Moderna Production Cost, $d_{Mod}$ (\$)	&	31.96	&	31.96	&	31.96	&	31.96	&	23.44	 \\
\hline
\hline
Private Sector Equilibrium Price, $p^{\text{priv}}$ (\$)	&	24.31	&	24.31	&	23.46	&	23.46	&	23.46	 \\
Private Sector Equilibrium Quantity, $q^{\text{priv}}$ (M)	&	55.6	&	55.6	&	53.6	&	53.6	&	53.6	 \\
\hline
Pfizer Public Sector Price, $p^{\text{pub}}_{Pf}$ (\$)	&	40.67	&	40.35	&	38.65	&	38.65	&	38.00	 \\
Pfizer Public Sector Quantity, $q^{\text{pub}}_{Pf}$ (M)	&	75.7	&	78.6	&	66.7	&	66.7	&	79.7	 \\
Pfizer Profit Realized (\$M)	&	234.0	&	234.0	&	\textbf{1015.9}	&	\textbf{1015.9}	&	25.7	 \\
\hline
Moderna Public Sector Price, $p^{\text{pub}}_{Mod}$ (\$)	&	53.66	&	54.92	&	47.15	&	47.15	&	53.86	 \\
Moderna Public Sector Quantity, $q^{\text{pub}}_{Mod}$ (M)	&	23.8	&	20.3	&	32.7	&	32.7	&	16.3	 \\
Moderna Profit Realized (\$M)	&	\textbf{90.2}	&	41.4	&	41.4	&	41.4	&	496.0	 \\
\hline

\end{tabu}
}
\end{center}
\caption{Scenarios in which public sector prices fall in the speculated ranges of $\$34-44$ per two doses for Pfizer and $\$45-55$ per two doses for Moderna.  Bold font indicates realized profit exceeding target profit.}
\label{T:rightprices}
\end{table}




\section{Conclusion} \label{s:conclusion}

In summary, we have adapted the model of \cite{CummingsEtAl2020} to the anticipated COVID-19 vaccine market duopoly between Pfizer-BioNTech and Moderna.  This paper extends the original model by incorporating production and distribution costs, anticipated to be substantial for  mRNA vaccines requiring cold storage.  Using publicly available information about the two companies and related vaccine markets, we use the model to recommend public sector prices to be negotiated by the CDC under a variety of parameter combinations.  Prices yielded by the model are realistic, averaging \$34-\$35 per two-dose regimen.  Moreover, several trends emerge: 
\begin{enumerate}
\item The CDC can negotiate prices with the manufacturers that keep public sector prices as low as possible while meeting demand and ensuring each manufacturer earns a target profit, even when production costs are high.
\item We can expect the public sector costs to decrease over time as the cold storage requirements of the vaccines loosen.
\item If the vaccines are differentiated through factors such as delivery method and adverse effects, we can expect disparate prices between the two manufacturers. 
\item Prices speculated in the media are observed as valid outputs of the model, indicating that such prices could satisfy the public sector demand, achieve the manufacturers' target profits, and minimize government costs.
\end{enumerate}

As better information becomes available about the COVID-19 vaccines, the model can be fine-tuned.  Moreover, while this paper has focused on Pfizer-BioNTech and Moderna, the two frontrunners in the COVID-19 vaccine race, there are other manufacturers currently developing COVID-19 vaccines intended for the U.S. market, including several that, as of this writing, are in clinical trials (\cite{nytvt}).  The original game theoretic model of \cite{bj} applies more generally to oligopolies, and the optimization model can also incorporate more than two manufacturers.  As the COVID-19 vaccine market evolves, the CDC and other decision-makers can use this model to guide pricing decisions.


%
%
%





\printbibliography


\vspace{5mm} 
{\it Acknowledgement.}
{The third author is supported by the National Science Foundation Graduate Research Fellowship under Grant No. 1122374. The opinion, findings, and conclusions or recommendations expressed in this material are those of the authors and do not necessarily reflect the opinions of the National Science Foundation.}

\end{document}